\documentclass[prb,showpacs,floatfix,twocolumn]{revtex4} 

\usepackage{epsfig,graphicx} 

\def\mno{Mn$_3$O$_4$} 

\begin{document} 

\title{Magnetodielectric coupling in \mno} 

\author{R. Tackett, G. Lawes} 

\affiliation{Department of Physics and Astronomy, Wayne State
University, Detroit, MI 48201} 

\author{B. C. Melot, M. Grossman, E. S. Toberer, R. Seshadri} 

\affiliation{Materials Department and Materials Research
Laboratory,  University of California, Santa Barbara, CA
93106} 

\date{\today} 

\begin{abstract} 

We have investigated the dielectric anomalies associated with
spin ordering transitions in the tetragonal spinel \mno\, using
thermodynamic, magnetic, and dielectric measurements. We 
find that two of the three magnetic ordering transitions in
\mno\, lead to decreases in the temperature dependent
dielectric  constant at zero applied field.  Applying a
magnetic field to the  polycrystalline sample leaves these two
dielectric anomalies  practically unchanged, but leads to an
increase in the dielectric  constant at the intermediate spin-ordering
transition. We discuss possible origins for this
magnetodielectric behavior in terms of spin-phonon coupling. Band structure
calculations suggest that in its ferrimagnetic state, Mn$_3$O$_4$
corresponds to a semiconductor with no orbital degeneracy due to 
strong Jahn-Teller distortion.

\end{abstract} 


\maketitle 

\section{Introduction}

Magnetodielectrics can be described as materials in which magnetic ordering
produces  dielectric anomalies, or alternately, materials in which the 
low-frequency dielectric constant is sensitive to an external magnetic 
field.\cite{newnham} These materials
have been studied for several decades, but the recent surge in activity in 
studying multiferroics\cite{Nicola_Science,Mathur_Nature} has prompted renewed 
interest in understanding the origins of the spin-charge coupling in these 
systems. While multiferroics are also magnetodielectric,  the converse
is frequently not true.  Nevertheless, characterizing the
shifts in the  dielectric constant induced by magnetic ordering
offers crucial insights into possible spin-charge coupling
mechanisms.  Additionally, the recognition that substantial
magnetodielectric coupling often arises in systems with
non-collinear magnetic structures\cite{SCO,Nagaosa,Mostovoy} 
offers the possibility that
dielectric spectroscopy may be a simple yet powerful tool for
identifying phase transitions among complex magnetic structures.

Many magnetodielectric systems have been investigated in the
past several years, including SeCuO$_3$,\cite{SCO}
EuTiO$_3$,\cite{takagi} BiMnO$_3$,\cite{tsuyoshi} 
CoCr$_2$O$_4$,\cite{CCO1,CCO2} and TmFeO$_3$,\cite{koo} in
addition to a very large set of studies on magnetocapacitive
coupling in multiferroics.\cite{Nicola_Science,Mathur_Nature}
Several models have been proposed
for explaining the observed dependence of the low-frequency
dielectric constant on spin structure and external magnetic
field. The simplest of these postulates a dielectric response which varies
as the square of the net
magnetization.\cite{takagi,tsuyoshi}
  These models fail to account for the large
dielectric shifts observed at antiferromagnetic transitions,
which can be qualitatively understood by considering a coupling
between the dielectric constant and the $q$-dependent
spin-spin correlation function.\cite{SCO}  It has
recently been recognized that magnetoresistive contributions in
inhomogeneous systems can also give rise to magnetocapacitive
effects.\cite{catalan,gavin_apl}           

\begin{figure} \smallskip \centering 
\epsfig{file=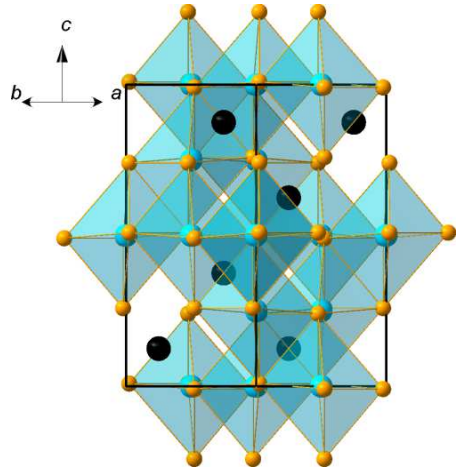, width=6cm} 
\caption{(Color online) Crystal structure of the tetragonal compound 
Mn$_3$O$_4$ derived from the cubic spinel through a strong Jahn-Teller 
distortion. The black spheres are tetrahedrally coordinated Mn$^{2+}$ on the
A site and the cyan spheres are the octahedrally coordinated Mn$^{3+}$ on the
B site. Orange spheres are O.}
\label{fig:struc}
\end{figure}

In the following, we discuss the synthesis and characterization
of the magnetodielectric spinel \mno\/ whose structure is displayed
in figure\,\ref{fig:struc}. This system has
been studied previously by Suzuki \textit{et al.},\cite{katsufuji}
who found that \mno\/ displays a sharp drop in the dielectric constant at
the ferrimagnetic $T_C$ = 42\,K.  This feature was attributed to the 
orbital degree of freedom in this system. Since \mno\ is known to exhibit
several complex low temperature spin 
states\cite{jensen,kuriki,hastings,tomiyasu} we conducted more
detailed  measurements of the dielectric constant at each of
the magnetic transitions in order to elucidate how specific
spin structures yield different magnetocapacitive couplings. 
At the onset of long  range magnetic order in \mno\, below 42\,K, 
the A-site Mn$^{2+}$ spins order ferromagnetically along [010],
while the B-site Mn$^{3+}$ spins order along [001]. Below 39\,K 
these B-site spins order in a spiral structure with an incommensurate
propagation vector along [010] and at 33\,K these  Mn$^{3+}$
ions exhibit a more complex order with the 16 ions in the
magnetic unit cell having a net moment antiparallel to the
Mn$^{2+}$ spin direction.\cite{jensen} 

\section{Experimental and computational details}

Well-sintered, brown pellets of Mn$_3$O$_4$ were prepared from
the  oxalate MnC$_2$O$_4$$\cdot$2H$_2$O by decomposing at
650$^\circ$C for  1\,h, pelletizing, and heating to
1200$^\circ$C for 12\,h following which the pellets were
rapidly cooled to the room temperature. Rapid cooling was required
in order to avoid producing Mn$_2$O$_3$ as an impurity phase. X-ray 
diffraction patterns were obtained using Cu-K$\alpha$ radiation on a Philips 
X'PERT MPD diffractometer operated at 45\,kV and 40\,mA, and 
subject to Rietveld refinement using the \textsc{xnd} Rietveld
code.\cite{xnd}

The DC magnetization of \mno\ was measured  a Quantum Design
SQUID magnetometer as a function of temperature and field. We
measured the specific heat and AC susceptibility using standard
options on a Quantum Design PPMS, which was also used to
regulate temperature and  magnetic field for dielectric
measurements.  For specific heat measurements, approximately 30\,mg
of the powder sample was cold pressed into a solid pellet,
which was found to have a small internal thermal time
constant.  To measure the  dielectric properties, we pressed a
circular pellet from approximately 50\,mg of sample, then
fashioned parallel plate electrodes using conducting silver
epoxy. The dielectric measurements were  done at an excitation
frequency of 30\,kHz with a drive of 1\,V.  There was no
significant frequency or bias dependence to the measurements. 

The electronic structure of Mn$_3$O$_4$ was 
calculated using density functional methods. For
these calculations, we used the linear muffin  tin orbital
(LMTO) method within the generalized gradient approximation
(GGA), as implemented in the \textsc{stuttgart tb-lmto-asa}
program.\cite{lmto}  Starting structures for LMTO calculations
were obtained from experimental  Rietveld refinements. 262
irreducible $k$-points were employed within the irreducible
part of the Brillouin zone.

\section{Results and discussion}

\begin{figure} \smallskip \centering
\epsfig{file=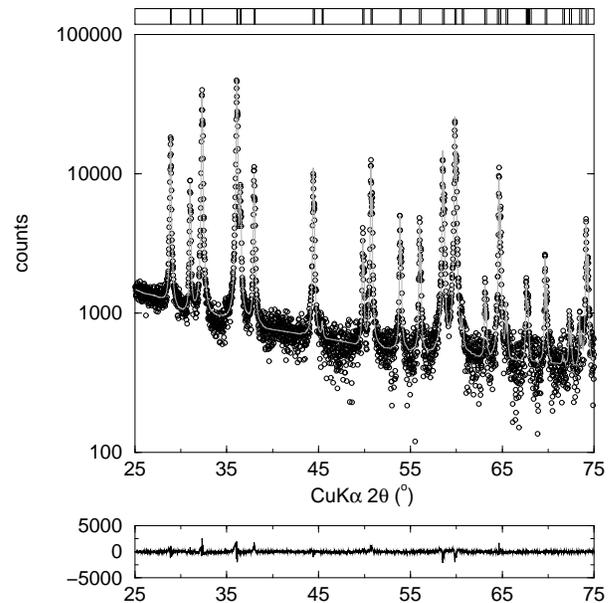, width=8cm}   \caption{X-ray diffraction
Rietveld analysis of Mn$_3$O$_4$  showing data as small circles
and fits as grey lines. The vertical lines at the top of the
figure indicate expected peak positions. Rietveld analysis gave 
$a$\,=\,5.7631$_1$\,\AA\/ and $c$\,=\,9.4712$_3$\,\AA. The position of the 
O atom was determined to be $(0,0.4700_5,0.2563_3)$. Subscripts
indicate experimental uncertainties in the last decimal place.} 
\label{fig:xray} 
\end{figure}

\mno\, is a tetragonally distorted spinel, as illustrated in
figure\,\ref{fig:struc}, with space group $I4_1/amd$ (No. 141).  
The polycrystalline samples used in this study were monophasic 
as revealed by X-ray diffraction Rietveld analysis shown in 
figure\,\ref{fig:xray}. In this structure,
Mn$^{2+}$ ions are located at the tetrahedral site at $(0,\frac14,\frac78)$, 
and Mn$^{3+}$ ions at the octahedral site$(0,\frac12,\frac12)$. 
O is at the general position $(0,y,z)$, More structural details are
provided in the caption of figure\,\ref{fig:struc}.
Suzuki \textit{et al.}\cite{katsufuji} have suggested that the
$e_g$ orbital at the Mn$^{3+}$ site is partially occupied, 
leading to an orbital degree of freedom. However, it is known that 
on cooling Mn$_3$O$_4$ undergoes a transition from a cubic spinel to 
Jahn-Teller distorted tetragonal structure at 
1433\,K.\cite{mn3o4_transition} As a consequence of this Jahn-Teller
structural distortion on Mn$^{3+}$, the electronic structure of tetragonal 
Mn$_3$O$_4$ as calculated here using density functional theory, 
and previously using Hartree-Fock theory\cite{Mn3O4_HF} 
suggests no orbital degeneracy.

\begin{figure}
\smallskip \centering
\epsfig{file=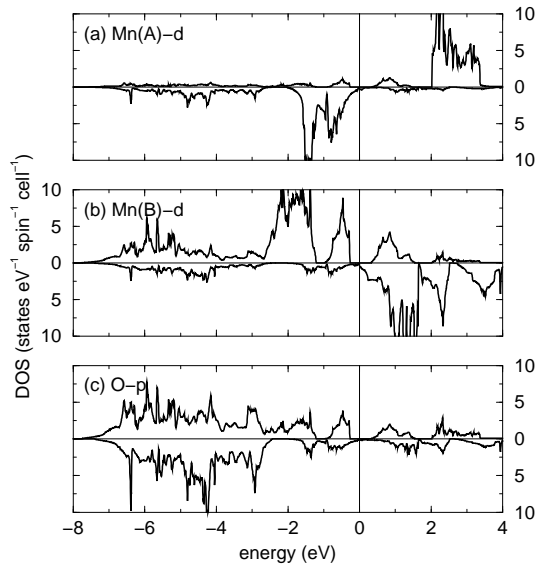, width=7cm}  \caption{Projected LMTO-GGA
Densities of state of semiconducting, N\'eel-ferrimagnetic Mn$_3$O$_4$
plotted in the two spin directions.  (a) Tetrahedral $d^5$ 
Mn$^{2+}$ on the A site showing filled minority $e$ and $t_2$ states, and
empty majority $e$ and $t_2$ states. (b) Octahedral Mn$^{3+}$ on the B site
with a Jahn-Teller distortion resulting in a gap within filled
and empty $e_g$ states: majority $d_{z^2}^1$ is separated from
majority  $d_{x^2-y^2}^0$. All minority states are empty. (c)
The O $p$ states. The origin on the energy axis is the top of the valence
band.}
\label{fig:dos}
\end{figure}

The LMTO-GGA density of states calculated for collinear ferrimagnetic 
Mn$_3$O$_4$, and projected on the different Mn atoms and on O are displayed 
in the panels of figure\,\ref{fig:dos}. The system is characterized by large
exchange splitting: near 4\,eV on the tetrahedral A site (occupied by
Mn$^{2+}$) and near 3\,eV on the octahedral B site (occupied by Mn$^{3+}$).
Both sites are completely spin-polarized, with the crystal field configurations
as indicated in the caption of the figure. The Jahn-Teller distortion on the
octahedral B site manifests clearly in the majority spin channel, which is
seen to comprise filled $t_{2g}^3$ states centered around $-$2\,eV, separated
from a filled $d_{z^2}^1$ state just below $E$ = 0. The empty majority 
$d_{x^2-y^2}$ states are found centered near 1\,eV. The minority Mn $d$ states 
at the B site are all empty and found starting at 0\,eV. The crystal field 
splitting on the B site is of the order of 4\,eV. O $p$ states are found
a little below the filled Mn $d$ states which is not surprising given the 
position of Mn in the middle of the first transition metal series. As would 
be expected for a 
fully spin-polarized ferrimagnet, the calculated magnetic moment per 
Mn$_3$O$_4$  formula unit is precisely 3\,$\mu_B$ corresponding to 
8 spins from two octahedral Mn$^{3+}$ ions, compensated by 5 spins from the
tetrahedral Mn$^{2+}$ ion.

\begin{figure} 
\smallskip \centering \epsfig{file=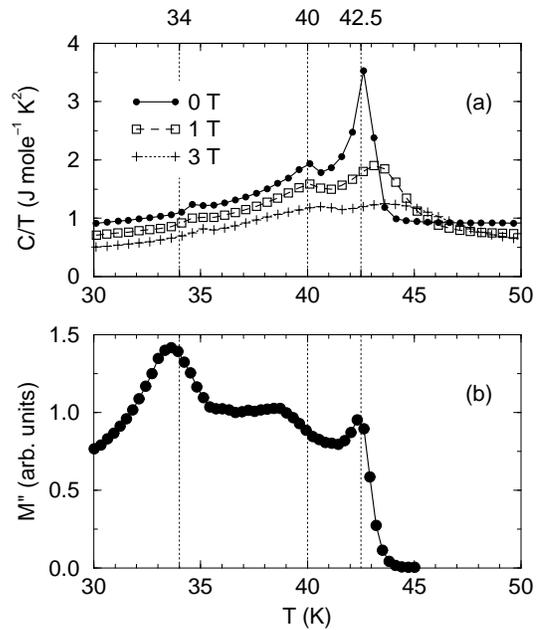, width=7cm}
\caption{(a) \mno\, ac susceptibility measured at $\omega/2\pi$ = 10\,kHz 
and zero field.  (b) \mno\, specific heat measured at $H$ = 0, $H$ = 1\,T, 
and $H$ = 3\,T. For clarity, the $H$ = 0\,T and $H$ = 1\,T data have 
been offset by 0.2 and 0.1\,J\,mole$^{-1}$\,K$^2$ respectively.}  
\label{fig:HCAC}
\end{figure}

In order to identify the numerous magnetic transitions in
\mno\, we first measured the specific heat capacity of this
sample, both at zero magnetic field and under the application
of a field.  These results are shown in panel (a) of 
figure\,\ref{fig:HCAC}.  There are three distinct peaks in heat
capacity at $H$ =0,  corresponding to magnetic phase transitions
at roughly $T$ = 34\,K, $T$ = 40\,K, and $T$ = 42\,K. 
Approximating the lattice
background to specific heat as a constant 0.5\,J\,mole$^{-1}$\,K$^2$ over
the the small temperature range of the peak, we find a total
spin entropy of  approximately 1.2\,$k_B$/\mno.  This is somewhat
smaller than previously reported values,\cite{kuriki} but
consistent with the suggestion that substantial geometrical
frustration reduces the entropy change from the
full value expected for complete spin order $\approx 5\, k_{B}$/\mno.  
These three peaks in \mno\, specific heat
correspond to the development of complex spin structures
investigated previously by neutron diffraction
measurements.\cite{jensen}  Applying a magnetic field produces
substantial broadening in the 42\,K peak, but does not
significantly shift the transition temperatures, at least for
fields up to $H$=3\,T.

\begin{figure} \smallskip \centering
\epsfig{file=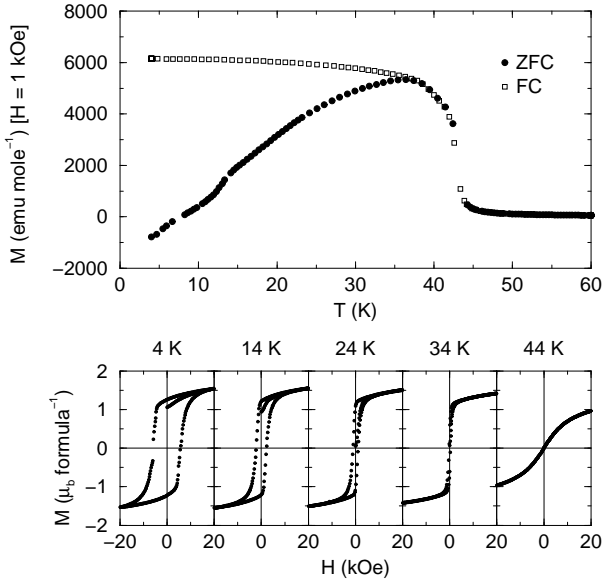, width=8cm} \caption{(a) Zero field cooled (ZFC) and 
field cooled (FC) magnetization curves for \mno\, measured in a field of 
1000\,Oe. (b) Magnetic hysteresis loops measured at temperatures from 
$T$ = 4\,K (far left) to $T$ = 44\,K 
(far right).} \label{fig:magn}
\end{figure}

We also see distinct features corresponding to these magnetic
transitions in \mno\, in ac susceptibility measurements.  The
imaginary component of the complex susceptibility measured at
10\,kHz is plotted in panel (b) of figure\,\ref{fig:HCAC}.  The
three peaks in the magnetic loss signal  correspond to the
three magnetic phase transitions observed the specific heat
data.  There is a substantial increase in the magnetic loss at
the onset of long-range magnetic order at 42\,K, which remains
high through the additional magnetic phase transitions at 40\,K
and  34\,K.  Small features are also observed in the real
component of  the complex magnetization (not shown), but these
these anomalies are  more difficult to discern against the
large signal from the 42\,K  transition.  The upper panel of 
figure\,\ref{fig:magn} plots the zero-field cooled and field
cooled temperature dependent DC magnetization under a measuring
field of 1000\,Oe, and the temperature dependent magnetic
hysteresis loops.  The transition to long-range order at 42\,K
is clearly visible, as is  a small anomaly at $T$ = 15\,K in the ZFC
curve, but the two additional magnetic phase transitions  at
$T$ = 34\,K and $T$ = 40\,K give no signal.  $M(H)$ loops plotted in the
lower  panel of figure\,\ref{fig:magn} show a net saturation
magnetization  corresponding to a moment of $\approx$1.5
$\mu_{B}$/formula unit, with  significant coercivity developing
only at temperatures below 34\,K. 

\begin{figure} \smallskip \centering
\epsfig{file=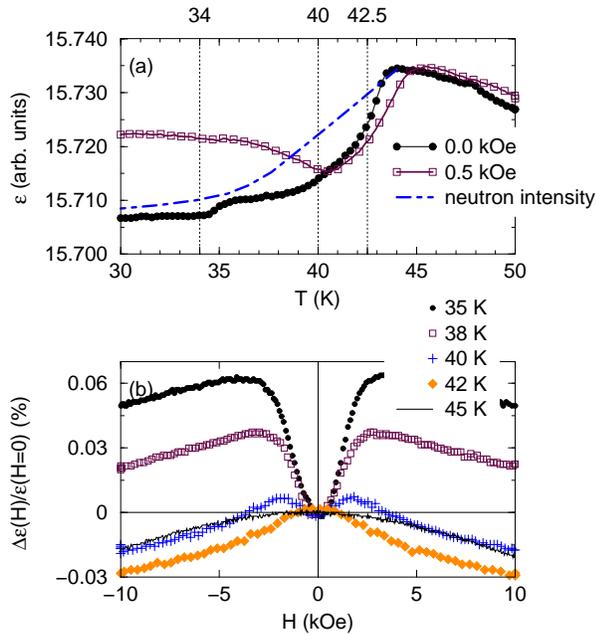, width=8cm}  \caption{(Color online) (a) Temperature
dependent dielectric constant of \mno\, at  $H$ = 0 (solid circles)
and $H$ = 5\,kOe (open squares). The broken line is the scaled neutron 
intensity of the (120) reflection taken from reference \onlinecite{jensen}. 
(b) Magnetic  field dependence of the dielectric constant of \mno\, at 
different  temperatures.}
\label{fig:mnodielec} \end{figure}

\begin{figure} 
\smallskip \centering \epsfig{file=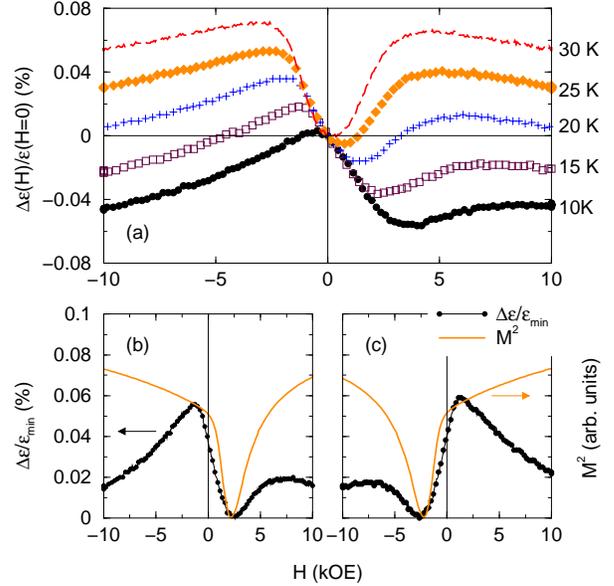, width=8cm}
\caption{(Color online) (a) Percentage change in \mno\, dielectric
constant (relative to $H$ = 0) as a function of increasing magnetic
field at fixed temperatures of 30\,K, 25\,K, 15\,K, and 10\,K (upper
trace to lower).  (b) Solid curve: percentage change in \mno\,
dielectric constant (relative to the minimum value). Dashed line:
square of the net magnetization.  Both are measured as a function
of increasing magnetic field measured at 15\,K. (c) Relative change
and net magnetization measured as a function of decreasing
magnetic field at 15\,K.}
\label{fig:loT} 
\end{figure}

The temperature dependent dielectric constant of \mno\, close
to  these magnetic ordering  transitions is
shown in figure\,\ref{fig:mnodielec}(a).  At zero applied
magnetic field, we see a  sharp decrease in the dielectric
constant at 42\,K,  coincident with the onset of long range
magnetic order  in \mno.  The dielectric constant decreases by
approximately  0.2\%, consistent with the magnitude of the drop
observed  previously in \mno.\cite{katsufuji}  In addition to the large
drop in dielectric constant at 42\,K, we  observe a much
smaller, but still distinct, negative shift of 0.015\% in  the
dielectric constant at the 34\,K magnetic transition.  A hint of
such  behavior is also visible in previous studies on
\mno.\cite{katsufuji}   While both the  42\,K and 34\,K magnetic
transitions in \mno\, lead to clear dielectric  anomalies at
zero field, there is no suggestion of any  magnetodielectric
coupling associated with the 40\,K transition.  However, the
application of a modest magnetic field substantially  alters
the dielectric response of \mno.  The negative 
magnetodielectric shifts associated with the 42\,K and 34\,K
transitions  remain relatively unchanged, but the 40\,K magnetic
transition now  produces a significant {\it increase} in
dielectric constant. 

In order to probe the effects of an applied magnetic field on
the  dielectric response of \mno\, we measured the dielectric
constant at  fixed temperature as a function of applied field. 
These data are  plotted in 
figure\,\ref{fig:mnodielec}(b). In the  paramagnetic phase, there is
a very slight decrease in dielectric  constant with magnetic
field.  Just below the onset of long-range  magnetic order, the
shift in dielectric constant with applied field remains
negative, but  with a change in curvature.  However,
the development of the spiral magnetic structure below 40\,K 
leads to positive  magnetodielectric shifts at low magnetic
fields, followed by smaller  decreases 
 at higher fields.  The increase in  dielectric
constant saturates at 
approximately 0.06\% in a field of 3\,kOe. 

We extended our measurements on the magnetic field dependence of
the dielectric constant to lower temperatures, below the 34\,K
transition.  As shown in figure\,\ref{fig:magn}, \mno\, develops magnetic
hysteresis at low temperatures, with the coercive field reaching
approximately 5\,kOe by $T$ = 10\,K.  Figure\,\ref{fig:loT}(a) plots the relative
shift in dielectric constant as a function of \emph{increasing}
magnetic field, measured at fixed temperatures.  At $T$ = 30\,K, there
is a slight asymmetry in the curve; this asymmetry increases
substantially as the temperature is reduced.  We attribute this
asymmetry to magnetic hysteresis in \mno\, at lower
temperatures.  Figures\,\ref{fig:loT}(b) and \, \ref{fig:loT}(c)
plot the change in dielectric constant relative to the minimum value, 
together with the square of the magnetization at $T$ = 15\,K, 
separately for increasing and decreasing magnetic fields.
While the relative change in dielectric
constant does not simply follow the square of the magnetization,
the minimum dielectric constant does occur at (approximately) the
coercive field, where $M$ = 0.  As we will discuss in the following,
we believe the dielectric shift in \mno\, is not determined by the
net magnetization, but rather a more complex spin-spin correlation
function for the magnetically ordered phase

\section{Discussion}

While both positive and negative magnetodielectric couplings
have been  observed previously,\cite{SCO} it is unusual for
both to be present in a single  system.  As such, \mno\,
provides an important test system for  understanding the
mechanisms underlying magnetodielectric effects.   Several
factors have been suggested to contribute to
magnetization-induced changes  in dielectric constant,
including spin-phonon coupling,\cite{SCO,brooks} electronic
structure,\cite{rai}, and orbital degrees of 
freedom.\cite{katsufuji}  Electromagnon scattering has also
been proposed to explain the increase in dielectric constant
observed at certain magnetic transitions in
multiferroics.\cite{electromagnon1,electromagnon2}
 In the following, we concentrate mainly
on discussing our experimental results in the  context of
possible spin-phonon coupling.

It has been suggested that spin-structure induced shifts in
phonon  frequencies can produce changes in the dielectric
constant through the  Lydanne-Sachs-Teller relation.\cite{SCO}
In this framework, the  optic  phonon frequencies, hence
dielectric constant, are coupled to the  spin-spin correlation
function for the magnetic structure, so
 the dielectric  constant should vary  with the neutron
peak intensity. The onset of  long range order in \mno, which
produces a significant drop in the  dielectric constant, is
associated with the growth of the (120)  neutron peak
intensity.\cite{jensen}  We have extracted the intensity  of
the (120) peak from reference \onlinecite{jensen} and plotted this
as a broken line in figure\,\ref{fig:mnodielec} after correcting for
a  small difference in transition temperature.  While the drop
in  dielectric constant in \mno\, is somewhat steeper than the
increase in  neutron intensity, the dielectric shift 
 approximately follows the intensity in the (120) peak and,
in particular, both level  off below 35\,K.  

While the qualitative agreement between the (120) neutron
intensity  and drop in the \mno\, dielectric constant below 42\,K are
consistent with the  spin-phonon model for magnetodielectric
coupling, the  detailed mechanism remains unclear.  As a first
step in  developing a microscopic model it is necessary to
identify which  specific phonon modes couple to the complex
spin structure.  The deformation produced by the
octahedral Jahn-Teller Mn$^{3+}$ ions gives rise to the tetragonal
distortion to $I4_{1}/amd$ symmetry in \mno.  This distortion
leads to 10 allowed Raman active modes,
$\Gamma=2A_{1g}+3B_{1g}+B_{2g}+4E_{g}$, rather than the 5 modes
in cubic spinels with $Fd\bar3m$  symmetry.\cite{lutz,malavasi}
Experimentally, only 5 Raman modes are  observed,\cite{lutz}
including a large amplitude $A_{1g}$ mode at 660\,cm$^{-1}$
attributed to the motion of oxygen ions inside the MnO$_{6}$
octahedra and associated with the cooperative Jahn-Teller
distortion.\cite{malavasi} Because magnetodielectric coupling
is often associated with non-collinear magnetic
structures,\cite{CCO1} it is tempting to  postulate that the
dielectric anomaly in \mno\, below 42\,K arises  because of
coupling between the complex spin structure exhibited by  the B site
Mn$^{3+}$ ions and vibrations of the MnO$_{6}$ octahedra.  
However, further  temperature dependent Raman studies are
needed to look for a shift or  increase in intensity of the 660\,cm$^{-1}$ 
peak at the 42\,K transition to test this  suggestion.   

As the (120) neutron scattering intensity varies smoothly below
42\,K,\cite{jensen} some additional spin-phonon coupling must
be responsible for the dielectric anomaly observed at 35\,K in
zero  field.  Below 40\,K, an additional neutron scattering
 reflection is observed
at  (1,1+$\tau$,0).   The propagation vector $q=1-\tau$
increases monotonically with  decreasing  temperature.  This peak
locks in at $q=1$ at the 34\,K magnetic  transition, which also
produces a sharp increase in intensity.\cite{jensen} 
We suppose that this low temperature  magnetic structure also
couples to an optic phonon mode in \mno, giving rise to the
dielectric  shift observed below 34\,K.  Between 34\,K and 40\,K,
$\tau$ is non-zero  and temperature dependent and the magnetic 
structure is incommensurate with the
lattice structure. 
The spin-phonon coupling coefficient is predicted to vary with 
the overlap between the $q$-dependent spin structure and phonon mode.\cite{SCO}
The incompatibility between spin and lattice structure could lead to a 
varying coupling coefficient, which averages to zero and gives no
net magnetodielectric coupling.
Only when the spin structure locks in at $q=1$ at 33\,K  
does the spin-phonon coupling constant become non-zero.  We
point out that incommensurate magnetic structures
can lead to  very dielectric anomalies associated with
transitions into  ferroelectrically ordered
phases.\cite{nvo,tmo1,tmo2}  This is distinctly different from
our observations in \mno, which seem to suggest that incommensurate 
magnetic structures do not produce sizeable dielectric anomalies.

It is rather more difficult to motivate
the positive shift in dielectric constant  below 40\,K in an
applied magnetic field, and positive magnetocapacitance in 
this temperature range (see figure\,\ref{fig:mnodielec}).   One
possibility is that even a modest magnetic field, below 3 kOe,
is  sufficient to lock-in some commensurate value for the
propagation  vector $q$ of the (110) structure, leading to
non-zero  magnetodielectric coupling.  Because the low temperature
dielectric constant shows hysteretic effects compatible with 
the magnetic hysteresis curves shown in figure\, \ref{fig:magn}, 
this (110) spin structure may also be expected to exhibit hysteresis.
Alternatively, some  additional mechanism for magnetodielectric
coupling beyond the  spin-phonon model considered here may be
relevant.  Additional studies  of the magnetic field dependence
of the \mno\, dielectric constant on  single crystal samples
may be required to clarify the origins of the  positive
magnetocapacitance observed between 34\,K and 40\,K at low 
fields. 

In summary, we have investigated the dielectric properties of
\mno\, in detail and found evidence for magnetodielectric 
coupling associated with specific magnetic structures.  An 
incommensurate magnetic transition in \mno\, produces no
dielectric  anomaly at zero magnetic field, but applying modest
fields produces an increase  in dielectric constant at this
transition. These magnetodielectric features can be 
qualitatively understood using a spin-phonon coupling model,
although  a more detailed understanding of the specific phonons
involved in the  coupling would be necessary to compare theory
and experiment in  detail.  

\acknowledgments Work at Wayne State was supported by the Donors of the 
American Chemical Society Petroleum Research Fund, and the Institute for
Manufacturing Research. Work at UCSB was supported by the National Science 
Foundation through a Career Award DMR 0449354, and by the NSF Chemical Bonding 
Center CHE 0434567. We also profited from facilities supported by the 
NSF MRSEC program DMR 0520415 at UCSB.

\clearpage

\end{document}